# New Generalized Morse-Like Potential for Studying the Atomic Interaction in Diatomic Molecules

C. M. Ekpo[1], Ephraim P. Inyang[2], Okoi P. O[2] and T. O. Magu[3], E. P. Agbo[1], K O Okorie[4], Etido P. Inyang[2],

[1]Department of Physics, Cross River University of Technology,Calabar, Nigeria

[2]Department of Physics, University of Calabar, Nigeria.

[3]Department of Pure and Applied Chemistry, University of Calabar, Calabar, Nigeria

[4]Department of Mathematics and Computer Science, Ebonyi State University, Abakaliki, Nigeria

## ABSTRACT

In this study, we obtain the approximate analytical solutions of the radial Schrodinger equation for the New Generalized Morse-Like Potential in arbitrary dimensions by using the Nikiforov Uvarov Method. Energy eigenvalues and corresponding eigenfunction are obtain analytically. The rotational-vibrational energy eigenvalues for some diatomic molecules are computed with the aid of some spectroscopic parameters. The energy equation for some potentials such as Attarctive Radial and Deng-Fan potentials have also been obtained by varying some potential parameters. Our results excellently agree with the already existing literature.

## 1 INTRODUCTION

The study of the bound state process is excruciatingly important when trying to understand molecular spectrum of diatomic molecules since the beginning of quantum mechanics [1,2]. The solutions of the Schrödinger equation with different potential models of interest have been employed to give insights, explanations and predictions into the behaviour of some class of molecules [3-4].

Quite recently, numerous researchers have proffered solutions to the Schrodinger equation for some potential of interest [5]. The exact or approximate solutions of these equations with the central potential play an important role in quantum mechanics [6-10].

The analytical solution of the Schrodinger equation with $l = 0$ and $l \neq 0$ for some potentials has been addressed by many researchers in nonrelativistic quantum mechanics and relativistic quantum mechanics for bound and scattering States[11-15]. Some of these potentials include, Shifted Deng-Fan potential[16], Generalised Morse potential [17], Hyperbolic potential [18], Hellmann Potential Hellmann−Frost−Musulin potential [19], Modified Tietz-Hua potential [20], and the Eckart

Manning Rosen Potential [21]. Also, several methods have been employed to obtain the solutions of the nonrelativistic wave equations with some potential models of interest, some these methods includes ; the Factorization Method [22], Formula Method[23], Modified Factorisation Method [7,8], Supersymmetry Quantum Mechanics (SUSYQM) [24-26], Nikiforov-Uvarov method (NU) [27] Asymptotic Iteration Method (AIM) [28,29], , Exact Quantization Rule [30],Proper Quantisation Rule[31] etc.

The potential under consideration (New Generalized Morse-Like Potential) proposed by[32] is given as;

$$V(r) = D_e \left[1 - \left(\frac{A+Be^{-\alpha r}}{C+De^{-\alpha r}}\right)^2\right] \quad (1)$$

which can be rewritten as;

$$V(r) = D_e - \frac{V_0}{\left(1+\frac{D}{C}e^{-\alpha r}\right)^2} - \frac{V_1 e^{-\alpha r}}{\left(1+\frac{D}{C}e^{-\alpha r}\right)^2} - \frac{V_2 e^{-2\alpha r}}{\left(1+\frac{D}{C}e^{-\alpha r}\right)^2} \quad (2)$$

Where $V_0 = D_e {A^2}/{C^2}$ $V_1 = D_e {2AB}/{C^2}$ $V_2 = D_e {B^2}/{C^2}$, $\quad (3)$

A, B, C, D, are constant coefficients and the term in the bracket is the Mobius square potential[14], $D_e$ is the dissociation energy and $\alpha$ is the screening parameter.
In the literature, this New Generalized Morse-Like Potential has been used to obtain the approximate bound state solutions of the radial part of the Dirac equation under spin symmetry, the eigenvalues and the corresponding wave function were obtained respectively via the Nikiforov-Uvarov method [32]. Isonguyo *et al.,*2014 also solved the Klein Gordon equation with this potential using SUSYQM formalism. The energy eigenvalues and the corresponding wave function were also obtained in a closed form for arbitrary $l$ state[33].

The potential understudy has not received great attention ever since it was proposed. The relativistic treatment for this potential has been carried out but to the best of our knowledge, none has been conducted in the Non-relativistic regime in the available literature. In view of this, we seek to carry out the Non Relativisic treatment for this potential as an extension to the works of Ikot *et al*,2013[32] and Isonguyo *et al*,2014[33] and also apply this to study some diatomic molecules.

The structure of our presentation is as follows. In the next section we give the basic nitty-gritty of the Nikiforov-Uvarov method and all necessary formulas required for our calculations. In Section 3, we solve the radial Schrödinger equation for the New Generalised Morse-like Potential and also obtain the rotational-vibrational energy spectrum for some classes of diatomic molecules ($H_2$, $LiH$, $CO$, $HCl$, $TiH$, $ScN$, $CuLi$, $CrH$, $NiC$ and $TiC$). Thereafter, we discuss our results in Section 4. Finally, we discuss special cases of the potential under consideration and conclusions are presented in Sections 5 and 6 respectively.

## 2 Review of the Nikiforov-Uvarov method

The Nikiforov-Uvarov (NU) method [34] is based on solving the hypergeometric-type second-order differential equations by means of the special orthogonal functions [35]. The main equation which is closely associated with the method is given in the following form [36]

$$\psi''(s) + \frac{\tilde{\tau}(s)}{\sigma(s)}\psi'(s) + \frac{\tilde{\sigma}(s)}{\sigma^2(s)}\psi(s) = 0 \qquad (4)$$

Where $\sigma(s)$ and $\tilde{\sigma}(s)$ are polynomials at most second-degree, $\tilde{\tau}(s)$ is a first-degree polynomial and $\psi(s)$ is a function of the hypergeometric-type.

The exact solution of Eq. (4) can be obtained by using the transformation

$$\psi(s) = \phi(s)y(s) \qquad (5)$$

This transformation reduces Eq. (4) into a hypergeometric-type equation of the form

$$\sigma(s)y''(s) + \tau(s)y'(s) + \lambda y(s) = 0 \qquad (6)$$

The function $\phi(s)$ can be defined as the logarithm derivative

$$\frac{\phi'(s)}{\phi(s)} = \frac{\pi(s)}{\sigma(s)} \qquad (7)$$

where $\pi(s) = \frac{1}{2}[\tau(s) - \tilde{\tau}(s)]$ \qquad (8)

with $\pi(s)$ being at most a first-degree polynomial. The second $\psi(s)$ being $y_n(n)$ in Eq. (5), is the hypergeometric function with its polynomial solution given by Rodrigues relation[36-37]

$$y^{(n)}(s) = \frac{N_n}{\rho(s)}\frac{d^n}{ds^n}[\sigma^n \rho(s)] \qquad (9)$$

where, $N_n$ is the normalization constant and $\rho(s)$ is the weight function which must satisfy the condition

$$(\sigma(s)\rho(s))' = \sigma(s)\tau(s) \qquad (10)$$

$$\tau(s) = \tilde{\tau}(s) + 2\pi(s) \qquad (11)$$

It should be noted that the derivative of $\tau(s)$ with respect to $s$ should be negative. The eigenfunctions and eigenvalues can be obtained using the definition of the following function $\pi(s)$ and parameter $\lambda$, respectively:

$$\pi(s) = \frac{\sigma'(s) - \tilde{\tau}(s)}{2} \pm \sqrt{\left(\frac{\sigma'(s) - \tilde{\tau}(s)}{2}\right)^2 - \tilde{\sigma}(s) + k\sigma(s)} \qquad (12)$$

where $k = \lambda - \pi'(s)$     (13)

The value of $k$ can be obtained by setting the discriminant of the square root in Eq. (12) equal to zero. As such, the new eigenvalue equation can be given as

$$\lambda_n = -n\tau'(s) - \frac{n(n-1)}{2}\sigma''(s), n = 0,1,2,... \tag{14}$$

This method has been used extensively by numerous researchers to solve various second-order differential equations arising in quantum mechanics such as Schrodinger equation, Klein-Gordon equation, Duffin-Kemmar-Petiau equation, spinless-Salpeter equation, and Dirac equations [27]. The success and power of this method need not be overemphasized.

## 3 Bound-state solutions

The radial Schrodinger equation in arbitrary dimensions [14, 38] can be given as:

$$\frac{d^2 R_{nl}}{dr^2} + \frac{2\mu}{\hbar^2}\left[E_{nl} - V(r) - \frac{\hbar^2(\delta+2\ell-1)(\delta+2\ell-3)}{8\mu r^2}\right]R_{nl} = 0 \tag{15}$$

where $\mu$ is the reduced mass, $E_{nl}$ is the energy spectrum, $\hbar$ is the reduced Planck's constant and $n$ and $l$ are the radial and orbital angular momentum quantum numbers respectively (or vibration-rotation quantum number in quantum chemistry). Substituting Eq. (3) into Eq. (15) gives:

$$\frac{d^2 R_{n\ell}(r)}{dr^2} + \left[\frac{2\mu E_{nl}}{\hbar^2} - \frac{2\mu D_e}{\hbar^2} - \frac{2\mu V_0}{\hbar^2\left(1+\frac{D}{C}e^{-\alpha r}\right)^2} - \frac{2\mu V_1 e^{-\alpha r}}{\hbar^2\left(1+\frac{D}{C}e^{-\alpha r}\right)^2} - \frac{2\mu V_2 e^{-2\alpha r}}{\hbar^2\left(1+\frac{D}{C}e^{-\alpha r}\right)^2} - \frac{(\delta+2\ell-1)(\delta+2\ell-3)}{4r^2}\right]R_{n\ell}(r) = 0$$
(16)

The radial Schrödinger equation for this potential can be solved exactly for $l = 0$ (s-wave) but cannot be solved for this potential for $l \neq 0$. To obtain the solution for $l \neq 0$, we employ the following new improved approximation scheme to deal with the centrifugal term (Pekeris-type approximation scheme), near the minimum [39]: to deal with the centrifugal term, which is given as;

$$\frac{1}{r^2} \approx \alpha^2\left[d_0 + \frac{e^{-\alpha r}}{\left(1+\frac{D}{C}e^{-\alpha r}\right)^2}\right] \tag{17}$$

where $\alpha r \ll 1$ and the dimensionless parameter $d_0 = \frac{1}{12}$. It has been shown that this approximation scheme is better than the Greene and Aldrich [40] approximation scheme used by [32-33], as pointed out by[41]. It is worthy to note that when $d_0 = 0$, the approximation Eq. (17) reduce to the usual approximation used in literature [41-44].

Substituting the approximation Eq. (17) into Eq. (16), we obtain an equation of the form;

$$\frac{d^2R_{n\ell}(r)}{dr^2} + \left[\frac{2\mu(E_{n\ell}-D_e)}{\hbar^2} - \frac{2\mu V_0}{\hbar^2\left(1+\frac{D}{C}e^{-\alpha r}\right)^2} - \frac{2\mu V_1 e^{-\alpha r}}{\hbar^2\left(1+\frac{D}{C}e^{-\alpha r}\right)^2} - \frac{2\mu V_2 e^{-2\alpha r}}{\hbar^2\left(1+\frac{D}{C}e^{-\alpha r}\right)^2} - \frac{(\delta+2\ell-1)(\delta+2\ell-3)}{4}\left(d_0 + \frac{e^{-\alpha r}}{\left(1+\frac{D}{C}e^{-\alpha r}\right)^2}\right)\right]R_{n\ell}(r) = 0 \qquad (18)$$

Eq. (18) can be simplified by introducing the following dimensionless abbreviations

$$\begin{cases} \varepsilon_n = -\frac{2\mu(E_{n\ell}-D_e)}{\hbar^2\alpha^2} \\ \beta_0 = \frac{2\mu V_0}{\hbar^2\alpha^2} \\ \beta_1 = \frac{2\mu V_1}{\hbar^2\alpha^2} \\ \beta_2 = \frac{2\mu V_2}{\hbar^2\alpha^2} \\ \gamma = \frac{(\delta+2\ell-1)(\delta+2\ell-3)}{4} \end{cases} \qquad (19)$$

$$\frac{d^2R_{n\ell}(r)}{dr^2} + \frac{1}{\left(1+\frac{D}{C}e^{-\alpha r}\right)^2}\left[-\varepsilon_n\left(1+\frac{D}{C}e^{-\alpha r}\right)^2 + \beta_0 + \beta_1 e^{-\alpha r} + \beta_2 e^{-2\alpha r} - \gamma\alpha^2\left(d_0\left(1+\frac{D}{C}e^{-\alpha r}\right)^2 + e^{-\alpha r}\right)\right]R_{n\ell}(r) = 0 \qquad (20)$$

Using a transformation $s = e^{-\alpha r}$ so as to enable us apply the NU method as a solution of the hypergeometric type

$$\frac{d^2R_{n\ell}(r)}{dr^2} = \alpha^2 s^2 \frac{d^2R_{n\ell}(s)}{ds^2} + \alpha^2 s \frac{dR_{n\ell}(s)}{ds} \qquad (21)$$

We obtain the differential equation

$$\frac{d^2R_{n\ell}}{ds^2} + \frac{\left(1+\frac{D}{C}s\right)}{s\left(1+\frac{D}{C}s\right)}\frac{dR_{n\ell}}{ds} + \frac{1}{s^2\left(1+\frac{D}{C}s\right)^2}\left[-\left(\varepsilon_n\frac{D^2}{C^2} - \beta_2 + \gamma d_0\frac{D^2}{C^2}\right)s^2 + \left(-2\varepsilon_n\frac{D}{C} - \gamma d_0\frac{D}{C} + \beta_1 - \gamma\right)s - (\varepsilon_n + \gamma d_0 - \beta_0)\right]R_{n\ell}(s) = 0 \qquad (22)$$

Comparing Eq. (22) and Eq. (4), we have the following parameters

$$\begin{cases} \tilde{\tau}(s) = \left(1+\frac{D}{C}s\right) \\ \sigma(s) = s\left(1+\frac{D}{C}s\right) \\ \tilde{\sigma}(s) = -\left(\varepsilon_n\frac{D^2}{C^2} - \beta_2 + \gamma d_0\frac{D^2}{C^2}\right)s^2 + \left(-2\varepsilon_n\frac{D}{C} - \gamma d_0\frac{D}{C} + \beta_1 - \gamma\right)s - (\varepsilon_n + \gamma d_0 - \beta_0) \end{cases} \qquad (23)$$

Substituting these polynomials into Eq. (15), we get $\pi(s)$ to be

$$\pi(s) = -\frac{s}{2} \pm \sqrt{(a+kq)s^2 + (b+k)s + c} \qquad (24)$$

where

$$\begin{cases} a = \frac{q^2}{4} + (\varepsilon_n q^2 - \beta_2 + \gamma d_0 q^2) \\ b = -(-2\varepsilon_n q - \gamma d_0 q + \beta_1 - \gamma) \\ c = (\varepsilon_n + \gamma d_0 - \beta_0) \\ q = \frac{D}{C} \end{cases} \quad (25)$$

To find the constant $k$, the discriminant of the expression under the square root of Eq. (24) should be equal to zero. As such, we have that

$$k_\pm = (\beta_1 - \gamma - 2\beta_0 q) \pm 2\sqrt{\varepsilon_n + \gamma d_0 - \beta_0} \sqrt{\frac{q^2}{4} - \beta_0 q^2 + \beta_1 q - \beta_2 - \gamma q} \quad (26)$$

Substituting Eq. (26) into Eq. (24) yields

$$\pi = -\frac{qs}{2} \pm \left[ \left( q\sqrt{\varepsilon_n + \gamma d_0 - \beta_0} - \sqrt{\frac{q^2}{4} - \beta_0 q^2 + \beta_1 q - \beta_2 - \gamma q} \right) s + \sqrt{\varepsilon_n + \gamma d_0 - \beta_0} \right] \quad (27)$$

From the knowledge of NU method, we choose the expression $\pi(s)_-$ which the function $\tau(s)$ has a negative derivative. This is given by

$$k_- = (\beta_1 - \gamma - 2\beta_0 q) \pm 2\sqrt{\varepsilon_n + \gamma d_0 - \beta_0} \sqrt{\frac{q^2}{4} - \beta_0 q^2 + \beta_1 q - \beta_2 - \gamma q} \quad (28)$$

with $\tau(s)$ being obtained as

$$\tau(s) = 1 + 2qs - 2\left( q(\sqrt{\varepsilon_n + \gamma d_0 - \beta_0})s - \sqrt{\frac{q^2}{4} - \beta_0 q^2 + \beta_1 q - \beta_2 - \gamma q} \right) s - 2\sqrt{\varepsilon_n + \gamma d_0 - \beta_0} \quad (29)$$

Referring to Eq. (13), we define the constant $\lambda$ as

$$\lambda = (\beta_1 - \gamma - 2\beta_0 q) - 2\sqrt{\varepsilon_n + \gamma d_0 - \beta_0} \sqrt{\frac{q^2}{4} - \beta_0 q^2 + \beta_1 q - \beta_2 - \gamma q} - \frac{q}{2} - \left( q\sqrt{\varepsilon_n + \gamma d_0 - \beta_0} - \sqrt{\frac{q^2}{4} - \beta_0 q^2 + \beta_1 q - \beta_2 - \gamma q} \right) \quad (30)$$

Taking the derivative of eq. (29) w. r. t. s, we have;

$$\tau'(s) = 2q - 2\left( q(\sqrt{\varepsilon_n + \gamma d_0 - \beta_0})s - \sqrt{\frac{q^2}{4} - \beta_0 q^2 + \beta_1 q - \beta_2 - \gamma q} \right) < 0 \quad (31)$$

And also taking the derivative of $\sigma(s)$ w. r. t. s from Eq.(23), we have;

$$\sigma''(s) = 2q \tag{32}$$

Substituting Eqs. (29) and (32) into Eq. (14), we obtain;

$$\lambda_n = -n^2 q - nq + 2nq\left(\sqrt{\varepsilon_n + \gamma d_0 - \beta_0}\right) - 2n\sqrt{\tfrac{q^2}{4} - \beta_0 q^2 + \beta_1 q - \beta_2 - \gamma q} \tag{33}$$

By comparing Eqs. (30) and (33), the exact energy eigenvalue equation is obtained as

$$\varepsilon_n = -\gamma d_0 + \beta_0 + \frac{1}{4}\left[\frac{\left(n+\tfrac{1}{2}+\sqrt{\tfrac{1}{4}-\beta_0+\tfrac{\beta_1}{q}-\tfrac{\beta_2}{q^2}-\tfrac{\gamma}{q}}\right)^2 - \beta_0 + \tfrac{\beta_2}{q^2}}{\left(n+\tfrac{1}{2}+\sqrt{\tfrac{1}{4}-\beta_0+\tfrac{\beta_1}{q}-\tfrac{\beta_2}{q^2}-\tfrac{\gamma}{q}}\right)}\right]^2 \tag{34}$$

Substituting Eqs. (23) into Eq. (34) yields the energy eigenvalue equation of the New Generalised Morse Potential in the form

$$E_{n\ell} = D_e - V_0 + \frac{\hbar^2\alpha^2 d_0(d+2\ell-1)(d+2\ell-3)}{8\mu} - \frac{\hbar^2\alpha^2}{8\mu}\left[\frac{\left(n+\tfrac{1}{2}+\sqrt{\tfrac{1}{4}+\zeta-\tfrac{(\delta+2\ell-1)(\delta+2\ell-3)}{4q}}\right)^2 - \tfrac{2\mu V_0}{\hbar^2\alpha^2} + \tfrac{2\mu V_2}{\hbar^2\alpha^2 q^2}}{\left(n+\tfrac{1}{2}+\sqrt{\tfrac{1}{4}+\zeta-\tfrac{(\delta+2\ell-1)(\delta+2\ell-3)}{4q}}\right)}\right]^2 \tag{35}$$

$$\zeta = -\frac{2\mu V_0}{\hbar^2\alpha^2} + \frac{2\mu V_1}{\hbar^2\alpha^2 q} - \frac{2\mu V_2}{\hbar^2\alpha^2 q^2} \tag{36}$$

The corresponding wave functions can be evaluated by substituting $\pi(s)\_$ and $\sigma(s)$ from Eq. (27) and Eq. (23) respectively into Eq. (7) and solving the first order differential equation. This gives

$$\Phi(s) = s^{-\sqrt{\varepsilon_n+\gamma d_0-\beta_0}}(1+qs)^{\left(\tfrac{q}{2}+\sqrt{\tfrac{q^2}{4}-\beta_0 q^2+\beta_1 q-\beta_2-\gamma q}\right)} \tag{37}$$

The weight function $\rho(s)$ from Eq. (10) can be obtained as

$$\rho(s) = s^{-2\sqrt{\varepsilon_n+\gamma d_0-\beta_0}}(1+qs)^{2\sqrt{\tfrac{q^2}{4}-\beta_0 q^2+\beta_1 q-\beta_2-\gamma q}} \tag{38}$$

From the Rodrigues relation of Eq. (9), we obtain

$$y_n(s) \equiv N_{n,l} P_n^{\left(-2\sqrt{\varepsilon_n+\gamma d_0-\beta_0},\, 2\sqrt{\tfrac{q^2}{4}-\beta_0 q^2+\beta_1 q-\beta_2-\gamma q}\right)}(1+2qs) \tag{39}$$

where $P_n^{(\theta,\vartheta)}$ is the Jacobi Polynomial.

Substituting $\Phi(s)$ and $y_n(s)$ from Eq. (38) and Eq. (40) respectively into Eq. (5), we obtain

$$\Psi_n(s) = N_{n,l} s^{-\sqrt{\varepsilon_n+\gamma d_0-\beta_0}}(1+qs)^{\left(\tfrac{q}{2}+\sqrt{\tfrac{q^2}{4}-\beta_0 q^2+\beta_1 q-\beta_2-\gamma q}\right)} P_n^{\left(-2\sqrt{\varepsilon_n+\gamma d_0-\beta_0},\, 2\sqrt{\tfrac{q^2}{4}-\beta_0 q^2+\beta_1 q-\beta_2-\gamma q}\right)}(1+2qs) \tag{40}$$

## 4 Results and Conclusion

In this study, we attempt to study the atomic interaction in diatomic molecules using the New Generalised Morse-like Potential. The bound state solution of this potential has been found in an arbitrary dimensions via NU method.

By using known spectroscopic values in Table 1, we obtained the energy states of some selected diatomic molecules for various vibrational $n$ and rotational $\ell$ angular momentum in 3D, as shown in Table 2. These diatomic molecules includes; $H_2$, LiH, CO, HCl, TiH, ScN, CuLi, CrH, NiC and TiC. The spectroscopic parameters were taken from the work of Oyewumi et al. [34], Hamzavi et al [37] and Berkdemir et al[45]. It is worthy to note that we applied the following conversion $1\, amu = 931.494028\, MeV/c^2$, $1 cm^{-1} = 1.23985 \times 10^{-4}\, eV$ and $\hbar c = 1973.29\, eV\, Å$ throughout our numerical computation.

For a fixed value of the principal quantum number $n$, the energy spectrum decreases as angular momentum quantum number increases, this can be seen explicitly in Table 2 and Fig's 3 and 4. When $\alpha \to 0 Å^{-1}$ and $D_e \to 0 eV$ (or for very small values of these parameters), in this regime the New Generalised Morse-like Potential model behaves like the Attractive Radial and Deng-Fan potentials as shown in Refs[32,33]. When applied to study diatomic molecules, this potential looks like an inverted Generalised Morse potential as seen in Fig's 1 and 2.

## 5 Conclusion

In this paper, the Schrödinger equation with New Generalised Morse-like Potential has been solved for arbitrary dimensions using NU method. In order to show the accuracy of our result, we have calculated the eigenvalues numerically for arbitrary $\ell$ and $n$ state. The general expressions obtained for the energy eigenvalues (Eq.34) and wave functions can be easily reduced to the 3D space ($\delta = 3$) and for s-wave (i.e. $\ell = 0$ state). We have also studied two special cases by adjusting the potential parameters and we found that this result reduces to those given in Refs. [5, 34]. This study stands as basis for the description of the quantum aspects of diatomic molecules using the New Generalised Morse-like Potential. We obtained the energy spectra of different diatomic molecules.

**Table 1** Model parameters of the diatomic molecules studied in the present work.

| Molecules | $D_e(eV)$ | $\alpha(\text{Å}^{-1})$ | $\mu(a.m.u)$ |
|---|---|---|---|
| $H_2$ | 4.744984 | 1.9426 | 0.50391 |
| $LiH$ | 2.515283695 | 1.128 | 0.880122 |
| $CO$ | 11.22696 | 2.2994 | 6.860672 |
| $HCl$ | 4.61962 | 1.8677 | 0.980105 |
| $TiH$ | 2.05 | 1.32408 | 0.987371 |
| $ScN$ | 4.56 | 1.5068 | 10.68277 |
| $CuLi$ | 1.74 | 1.00818 | 6.259494 |
| $CrH$ | 2.13 | 1.52179 | 0.988976 |
| $NiC$ | 2.76 | 2.25297 | 9.974265 |
| $TiC$ | 2.66 | 1.5255 | 9.606079 |

**Table 2** ; Energy spectra of the New Generalised Morse-like Potential for ($H_2$, LiH , CO, HCl, TiH, ScN, CuLi, CrH, NiC and TiC) molecules for various n and rotational $\ell$ quantum numbers.

| $n$ | $\ell$ | $H_2$ | $LiH$ | $CO$ | $HCl$ | $TiH$ | $ScN$ | $CuLi$ | $CrH$ | $NiC$ | $TiC$ |
|---|---|---|---|---|---|---|---|---|---|---|---|
| 0 | 0 | -0.003913142 | -0.000755417 | -0.00040269 | -0.001859748 | -0.000927812 | -0.000111055 | -0.00008485 | -0.001223589 | -0.000265913 | -0.000126587 |
|   | 1 | -0.013043805 | -0.002518058 | -0.001342308 | -0.006199155 | -0.003092706 | -0.000370184 | -0.00028283 | -0.004078629 | -0.00088638 | -0.000421957 |
| 1 | 0 | -0.015652566 | -0.00302167 | -0.001610766 | -0.007438988 | -0.003711247 | -0.000444221 | -0.000339394 | -0.004894356 | -0.001063655 | -0.000506348 |
|   | 1 | -0.032609513 | -0.006295146 | -0.003355772 | -0.015497885 | -0.007731764 | -0.00092546 | -0.000707077 | -0.010196573 | -0.002215951 | -0.001054893 |
| 2 | 0 | -0.035218274 | -0.006798757 | -0.003624228 | -0.01673772 | -0.008350305 | -0.000999497 | -0.000763639 | -0.0110123 | -0.002393225 | -0.001139284 |
|   | 1 | -0.060001504 | -0.011583068 | -0.006174623 | -0.028516108 | -0.014226446 | -0.001702847 | -0.001301022 | -0.018761694 | -0.004077351 | -0.001941004 |
|   | 2 | -0.090002256 | -0.017374602 | -0.009261938 | -0.042774159 | -0.021339669 | -0.002554271 | -0.001951534 | -0.028142541 | -0.006116028 | -0.002911507 |
| 3 | 0 | -0.062610265 | -0.01208668 | -0.006443076 | -0.029755944 | -0.014844987 | -0.001776884 | -0.001357583 | -0.019577422 | -0.004254624 | -0.002025394 |
|   | 1 | -0.095219778 | -0.018381826 | -0.00979886 | -0.045253822 | -0.022576752 | -0.002702344 | -0.002064666 | -0.029773993 | -0.006470579 | -0.003080289 |
|   | 2 | -0.133046813 | -0.025684195 | -0.013691561 | -0.063231365 | -0.031545598 | -0.003775878 | -0.002884877 | -0.041602017 | -0.009041085 | -0.004303967 |
|   | 3 | -0.176091371 | -0.033993787 | -0.018121186 | -0.08368857 | -0.041751527 | -0.004997486 | -0.003818221 | -0.055061493 | -0.011966142 | -0.005696427 |
| 4 | 0 | -0.097828539 | -0.018885437 | -0.010067311 | -0.04649366 | -0.023195293 | -0.002776381 | -0.002121226 | -0.030589721 | -0.006647852 | -0.003164679 |
|   | 1 | -0.138264336 | -0.026691418 | -0.014228483 | -0.065711028 | -0.032782681 | -0.003923952 | -0.002998008 | -0.043233469 | -0.009395636 | -0.004472749 |
|   | 2 | -0.183917654 | -0.035504622 | -0.018926571 | -0.087408063 | -0.043607151 | -0.005219596 | -0.003987919 | -0.057508671 | -0.01249797 | -0.005949601 |
|   | 3 | -0.234788494 | -0.04532505 | -0.024161582 | -0.11158476 | -0.055668703 | -0.006663315 | -0.005090961 | -0.073415324 | -0.015954856 | -0.007595236 |
|   | 4 | -0.290876857 | -0.056152701 | -0.029933516 | -0.138241118 | -0.068967338 | -0.008255107 | -0.006307136 | -0.090953429 | -0.019766295 | -0.009409653 |
| 5 | 0 | -0.140873097 | -0.02719503 | -0.014496932 | -0.066950868 | -0.033401222 | -0.003997989 | -0.003054568 | -0.044049197 | -0.009572908 | -0.004557139 |
|   | 1 | -0.189135176 | -0.036511846 | -0.019463491 | -0.089887726 | -0.044844233 | -0.00536767 | -0.00410105 | -0.059140123 | -0.012852522 | -0.006118384 |
|   | 2 | -0.242614777 | -0.046835885 | -0.024966966 | -0.115304253 | -0.057524326 | -0.006885425 | -0.005260659 | -0.075862502 | -0.016486685 | -0.00784841 |
|   | 3 | -0.301311901 | -0.058167147 | -0.031007364 | -0.143200442 | -0.071441502 | -0.008551254 | -0.0065334 | -0.094216333 | -0.020475399 | -0.009747219 |
|   | 4 | -0.365226547 | -0.070505633 | -0.037584684 | -0.173576292 | -0.08659576 | -0.010365156 | -0.007919274 | -0.114201615 | -0.024818667 | -0.011814812 |
|   | 5 | -0.434358714 | -0.083851342 | -0.044698929 | -0.206431803 | -0.1029871 | -0.012327132 | -0.00941828 | -0.135818349 | -0.029516486 | -0.014051187 |

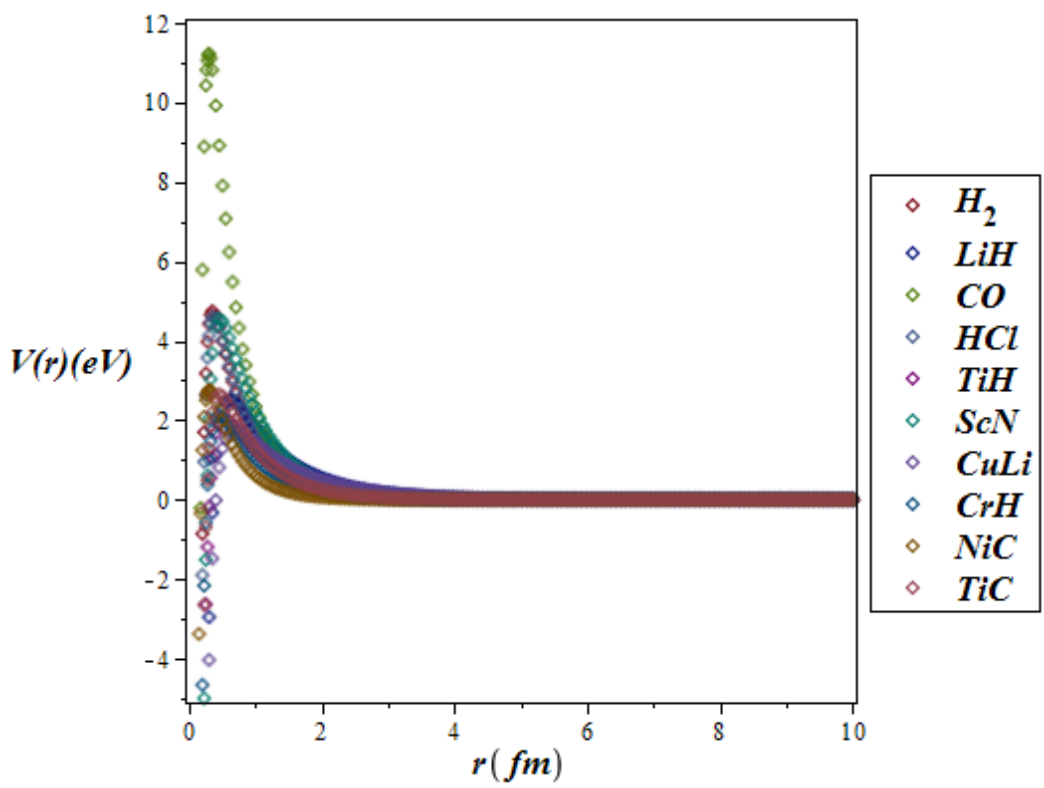

Figure 1; Shape of the New Generalised Morse-like Potential for different diatomic molecules($H_2$, LiH, CO, HCl, TiH, ScN, CuLi, CrH, NiC and TiC). $A = 1, B = -2, C = 1, D = -1$.

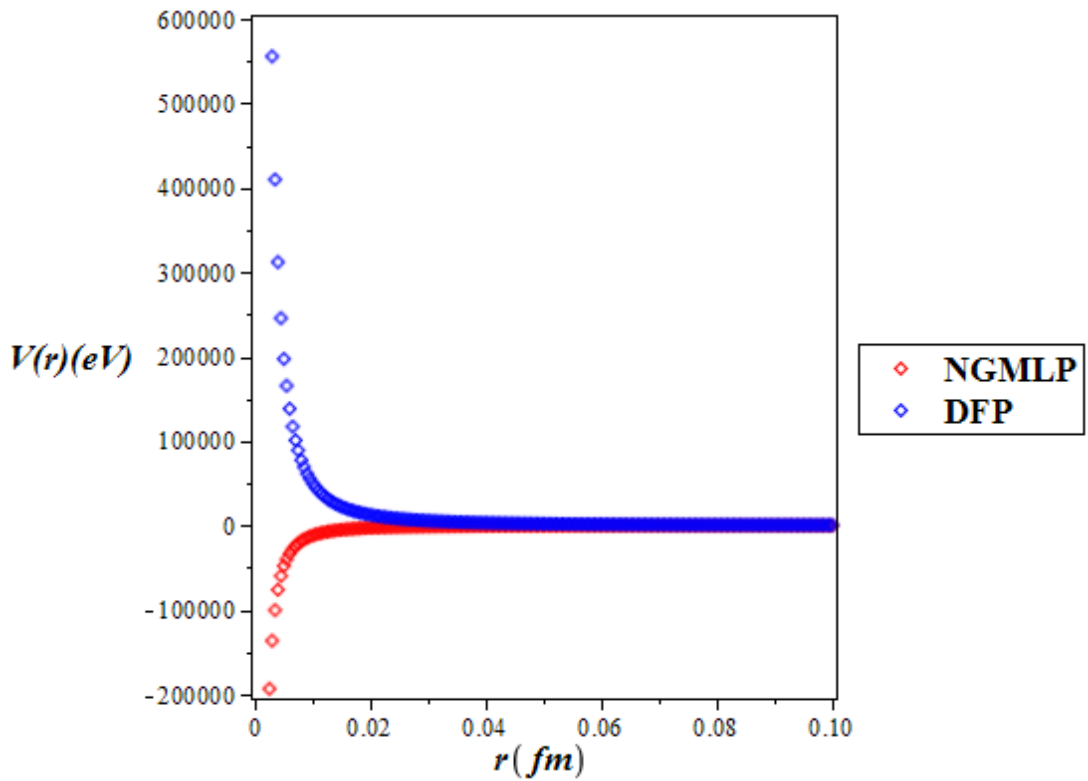

**Fig. 2;** Shape of the Deng-Fang and New Generalised Morse-like Potentials for $H_2$ diatomic molecule, with $A = 1, B = -2, C = 1, D = -1$

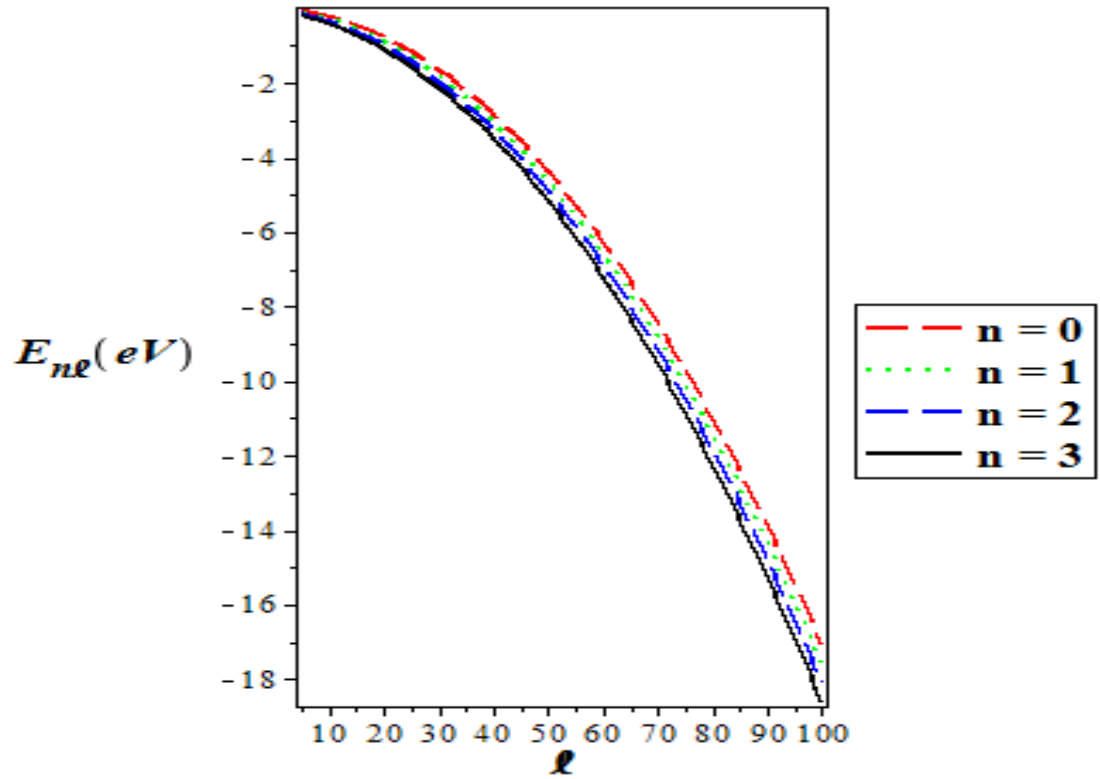

Figure 3; The variation of the energy spectrum for various values of $n$ as a function of $l$. We choose; $\hbar = 1\ 2\mu = 1, A = D = -1, D = B = 1, D_e = 1$ and $\alpha = 0.1$ in 3D

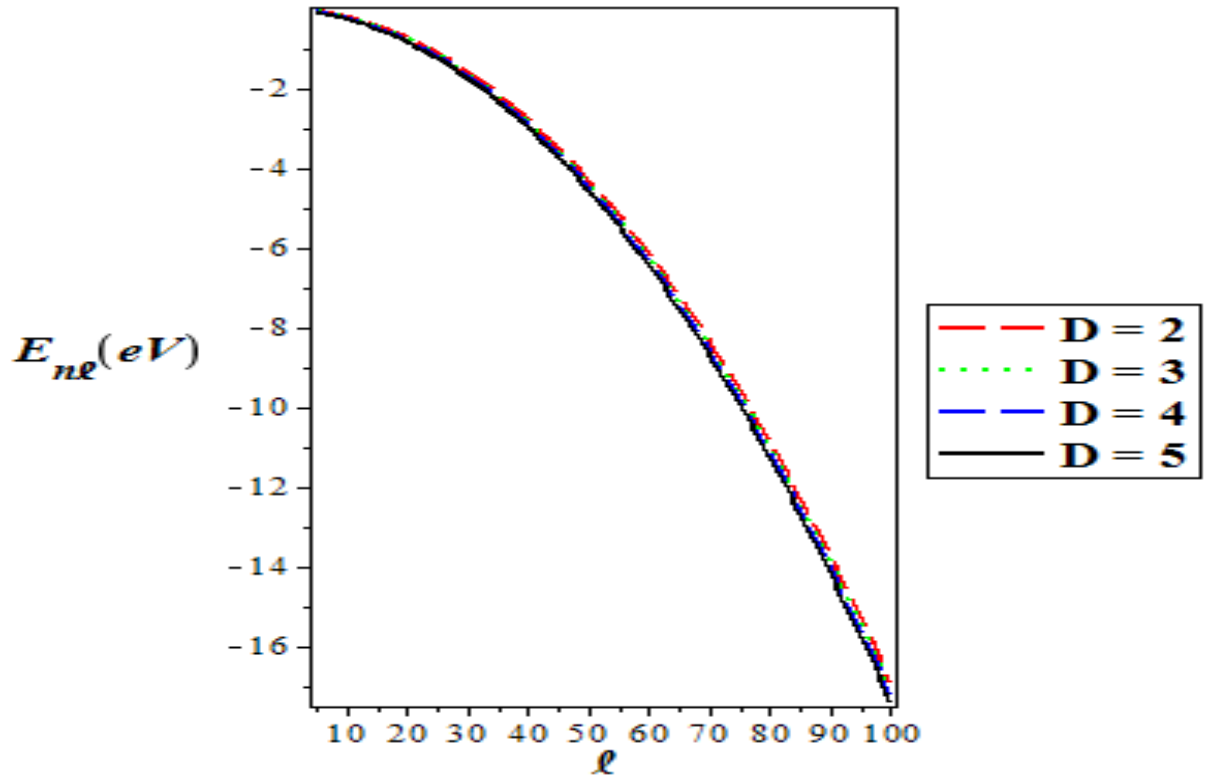

Figure 4; The variation of the Ground state energy spectrum for various values of dimension $D$ as a function of $l$. We choose; $\hbar = 1$ $2\mu = 1$, $A = D = -1$, $C = B = 1$, $D_e = 1$ and $\alpha = 0.1$